\documentclass[aps,prb,twocolumn,amsmath,amssymb,superscriptaddress,floatfix]{revtex4}
\pdfoutput=1

\usepackage{graphicx}
\usepackage{graphics}
\usepackage{bm}
\usepackage{amsfonts}
\usepackage{amssymb}
\usepackage{amsmath}
\usepackage{bbold}
\usepackage[usenames]{color}
\usepackage{colordvi}
\usepackage{units}
\usepackage{bbm}
\usepackage{booktabs}
\usepackage{array}
\usepackage{placeins}
\usepackage{epsfig}
\usepackage{lscape}
\usepackage{changes}
\usepackage{braket}
\usepackage{tabularx}
\newcolumntype{L}[1]{>{\raggedright\arraybackslash}p{#1}} % linksbündig mit Breitenangabe
\newcolumntype{C}[1]{>{\centering\arraybackslash}p{#1}} % zentriert mit Breitenangabe
\newcolumntype{R}[1]{>{\raggedleft\arraybackslash}p{#1}} % rechtsbündig mit Breitenangabe

\newcommand{\be}{\begin{equation}}
\newcommand{\ee}{\end{equation}}
\newcommand{\beqn}{\begin{eqnarray}}
\newcommand{\eeqn}{\end{eqnarray}}
\newcommand{\Tr}{\textnormal{Tr}}
\bibstyle{apsrev.bib}

\begin{document}

\title{Entanglement entropy of random partitioning}
\author{Gerg\H o Ro\'osz}
\email{roosz.gergo@wigner.mta.hu}
\affiliation{Institute of Theoretical Physics, Technische Universit\"at Dresden, 01062 Dresden, Germany}
\affiliation{Wigner Research Centre for Physics, Institute for Solid State Physics and Optics, H-1525 Budapest, P.O. Box 49, Hungary}
\author{Istv\'an A. Kov\'acs}
\email{istvan.kovacs@northwestern.edu}
\affiliation{Department of Physics and Astronomy, Northwestern University, Evanston, IL 60208-3112, USA}
\affiliation{Wigner Research Centre for Physics, Institute for Solid State Physics and Optics, H-1525 Budapest, P.O. Box 49, Hungary} 
\affiliation{Department of Network and Data Science, Central European University, Budapest, H-1051, Hungary}

\author{Ferenc Igl{\'o}i}
\email{igloi.ferenc@wigner.mta.hu}
\affiliation{Wigner Research Centre for Physics, Institute for Solid State Physics and Optics, H-1525 Budapest, P.O. Box 49, Hungary}
\affiliation{Institute of Theoretical Physics, Szeged University, H-6720 Szeged, Hungary}
\date{\today}

\begin{abstract}
	We study the entanglement entropy of random partitions in one- and two-dimensional critical fermionic systems. In an infinite system we consider a finite, connected (hypercubic) domain of linear extent $L$, the points of which with probability $p$ belong to the subsystem. The leading contribution to the average entanglement entropy is found to scale with the volume as $a(p) L^D$, where $a(p)$ is a non-universal function, to which there is a logarithmic correction term, $b(p)L^{D-1}\ln L$. In $1D$ the prefactor is given by $b(p)=\frac{c}{3} f(p)$, where $c$ is the central charge of the model and $f(p)$ is a universal function. In $2D$ the prefactor has a different functional form of $p$ below and above the percolation threshold.
\end{abstract}


\maketitle

\section{Introduction}
The entanglement properties of many-body quantum systems are subjects of recent intensive theoretical studies \cite{amico2008, calabrese2009, eisert2010, area_2}.
The entanglement between two partitions $A$ and $B$ of a system being in a pure state $| \Psi \rangle$ can be  measured by the entanglement entropy\cite{benett1996}: $S=- \textnormal{Tr}_B \rho_B \ln \rho_B=- \textnormal{Tr}_A \rho_A \ln \rho_A$. Here $\rho_A= \textnormal{Tr}_B \rho$ and $\rho_B= \textnormal{Tr}_A \rho$ are the reduced density matrices of the subsystems $A$ and $B$, respectively, and $\rho=| \Psi \rangle  \langle \Psi |$.

Most of the studies are restricted to bipartitions with a smooth, regular boundary between $A$ and $B$, for example in one dimension ($1D$) the subsystem $A$ contains the successive sites $i=1, \, 2, \, \dots \, L$, and $B$ is represented by the rest of the sites.
If the system is gapped the entanglement entropy generally satisfies the so-called area law \cite{eisert2010}: $S \sim L^{D-1}$. In one-dimensional critical systems, with algebraically decaying correlations, the area law is supplemented by a logarithmic correction, which for conformally invariant systems is given by \cite{holzhey1994, calabrese2004, corr-matrix-method, peschel-2003, jin2004, igloi-juhasz2008}
\begin{equation}
 S (L) \simeq \frac{c}{3}  \ln L + c_1
 \label{pure-1D-ent}
\end{equation}
where $c$ is the central charge of the conformal algebra. 

Multidimensional ($D>1$) free fermion systems satisfy an area law if the spectrum is gapped \cite{wolf2006} or the Fermi surface has high codimension \cite{li2006}. In the case of a sharp $D-1$ dimensional Fermi surface, there is a logarithmic correction to the are law\cite{wolf2006, farkas-zimboras-2007}, and the entanglement entropy is given by the following expression \cite{area_1, barthel2006, gioev-2006}
\begin{equation}
S(L)= \frac{L^{d-1}}{(2\pi)^{d-1}} \frac{\ln L}{12} \int \int | n_x \cdot n_k | d A_x d A_k \;,
\label{multidim-ent}
\end{equation}
where the integral is over a scaled version of the spatial and the Fermi surface, in such a way that the volume of the (scaled) Fermi sea is $1$, and
 $n_x$ and $n_k$ are unit normals to the real space boundary and the Fermi surface, respectively.

In disordered quantum spin chains the average entanglement entropy at the critical point has a logarithmic size dependence\cite{refael,Santachiara,Bonesteel,s=1,Laflo05,igloi-yu-cheng-ent,dyn06}, too, which can be calculated by the strong disorder RG method\cite{im}. In higher dimensional, critical random quantum systems there is an additive logarithmic correction due to corners, the prefactor of which is universal, i.e. independent of the form of disorder\cite{random_entr_d}.

If the subsystem $A$ is not a singly connected domain, much less (analytical) results are available. Here we mention that if $A$ and $B$ contain the sites of two sublattices the contact points between them scale with the volume of the system and so behaves the entanglement entropy, too\cite{fermion-and-spin-ent}. The entanglement entropy of irregular subsystems with non-continuous border is also subject of research in the recent years. General upper and lower bounds have been set for fractal boundary in real space and fractal like Fermi-surface in Ref. \cite{gioev-2006}. 
Fractal bipartition in the topologically ordered phase of the toric code with a magnetic field  was also investigated in Ref. \cite{fractal-boundary-real}. 

In the present paper we study the entanglement entropy when $A$ is elected by a random partition, which means that points of a given domain belong to $A$ with some probability $p$. This type of setting has already been used in Ref. \cite{random_partition}, where the low-lying part of the entanglement Hamiltonian of a random partition is calculated for the non-critical Kitaev-chain\cite{kitaev-model}. Here we consider critical fermionic models, hopping models in $1D$ and $2D$, as well as the critical Kitaev-chain. 

The rest of the paper is organised as follows. Models and the methods of calculations are presented in Sec.\ref{sec:models}.
Lower and upper bounds for the entanglement entropy are calculated in Sec.\ref{sec:bounds}, while small $p$ and small $1-p$ expansions are performed in Sec.\ref{sec:series-exp}. This is followed by extensive numerical calculations in Sec.\ref{sec:numeric}, for different values of the occupation probability $p$ and the linear size of the (hypercubic) domain, $L$. We discuss our results in Sec.\ref{sec:discussion} while detailed calculations are put to the Appendices.

\section{Models and methods}
\label{sec:models}

We consider fermionic hopping models with half filling defined by the Hamiltonian
\be
H=-t\sum_{\langle i,j \rangle} c_i^{\dagger} c_{j}\;,
\label{H_hopping}
\ee
in terms of the fermion creation, $c_i^{\dagger}$, and annihilation, $c_{j}$, operators at lattice sites $i$ and $j$, respectively and the summation runs over nearest neighbour lattice sites. The lattice is either an infinite chain ($1D$) or an infinite square lattice ($2D$), in the latter case the components of the positions are $i=(i_x,i_y)$ and $j=(j_x,j_y)$.

Having the two-point correlation function, $C(i,j)=\langle c^{\dagger}_i c_j\rangle$ for $i,j \in A$, we can calculate the entanglement entropy of the system as
\begin{align}
 S=& -\textnormal{Tr}_{A} \left[ C \ln C + (1-C) \ln(1-C)  \right]  \nonumber \\
  =& - \sum_{i=1}^{N_A} \left[ \zeta_i \ln \zeta_i + (1-\zeta_i) \ln(1-\zeta_i)  \right]=- \sum_{i=1}^{N_A} s(\zeta_i)\;, \label{S-eig}
\end{align}
where $N_A$ is the dimension of the correlation matrix (number of sites in the subsystem), and $\zeta_i$ are the eigenvalues of the correlation matrix. As discussed in the introduction we consider finite domains of linear extent, $L$, (subsequent points in $1D$ and a square in $2D$) the points of which belong to the subsystem $A$ with probability $p$. We have $2^L$ ($2^{L^2}$) different subsystems in $1D$ ($2D$), for which the entanglement entropy needs to be averaged, while the average value of $N_A$ is $pL$ ($pL^2$) in $1D$ ($2D$).

For the hopping model $C(i,i)=1/2$, whereas for $i \ne j$ we have
\be
C(i,j)=\frac{1} {\pi (i-j)} \sin  \frac{\pi (i-j)}{2}\;
\label{1D-corr-func}
\ee
in $1D$ and
\begin{equation}
C(i,j)= \begin{cases}
           0  \textnormal{, if  }  \quad (i_x-j_x)^2-(i_y-j_y)^2=0   \\
           \\
           \displaystyle{-\frac{(-1)^{i_x-j_x}-(-1)^{i_y-j_y}}{\pi^2 \left[ (i_x-j_x)^2 - (i_y-j_y)^2 \right]}}  \textnormal{, otherwise}   \\
          \end{cases}\;,
\label{2D-corr-func}
\end{equation}
in $2D$.

For the non-random partition (with $p=1$) the bipartite entanglement entropy in $1D$ is given by Eq.(\ref{pure-1D-ent}) with the central charge $c^{\textrm{hop}}=1$ and the constant is $c_1^{\textrm{hop}}=\ln(2)/3+(1+\gamma_E)/3-1/30 \approx 0.723$, where $\gamma_E$ is the Euler constant \cite{jin2004}.
In $2D$ for an $L \times L$ subsystem the prefactor of $L \ln L$ in Eq.(\ref{multidim-ent}) is given by $2/3$, which has been verified by numerical calculations \cite{weifei2006,barthel2006}.

Our second fermionic model is the critical Kitaev chain defined by the Hamiltonian\cite{kitaev-model}
\begin{align}
H_{Kit}= & -\sum_{l=-\infty}^{l=+\infty} \left[( c_{l+1}^{\dagger} c_l + c_l^{\dagger} c_{l+1} ) - (c_{l+1}^{\dagger} c_l^{\dagger} + c_l c_{l+1})\right. \nonumber \\
             & \left. +(c^{\dagger}_l c_l -1/2)\right] \;.
\label{kit-chain}
\end{align}
This model corresponds to the fermionic form of the critical quantum Ising chain, what is obtained after performing the standard Jordan-Wigner transformation\cite{pfeuty}.
The relevant correlation function for this model is
\begin{equation}
C(i,j)=\langle (c_i^{\dagger}-c_i)( c_j^{\dagger} + c_j )\rangle \;,
\label{kit-corr}
\end{equation}
given by\cite{igloi-yu-cheng-ent}
\begin{equation}
C(i,j)= \frac{2}{\pi} \frac{(-1)^{i-j}}{2(i-j)+1} \;.
\end{equation}
For the non-random partition (with $p=1$) the entanglement entropy of the Kitaev chain (or the critical quantum Ising chain) is again given by Eq.(\ref{pure-1D-ent}), with the central charge $c^{\textrm{Kit}}=1/2$ and the constant is $c_1^{\textrm{Kit}}=c_1^{\textrm{hop}}/2+c^{\textrm{Kit}}/3 \approx 0.528$. Note, however, that if the subsystem $A$ is not a single connected domain, as is the case for random partitions, then the entanglement entropy of the critical Kitaev chain and that of the critical quantum Ising chain is different, due to the non-local nature of the Jordan-Wigner transformation\cite{fermion-and-spin-ent}.

\section{Lower and upper bounds from particle number fluctuations}
\label{sec:bounds}
Following standard techniques\cite{peschel-viktor-review}, we can calculate a lower bound to the entanglement entropy by using the inequality
\begin{equation}
s(x) \ge 4 \ln (2)  x(1-x)
\label{ineq-S}
\end{equation}
where $s(x)$ is defined in Eq.(\ref{S-eig}). Then, for the entropy we obtain 
\begin{equation}
S \ge 4 \ln 2 \textnormal{Tr} \left[ C - C^2 \right]=4 \ln 2  \left[\langle N^2 \rangle - \langle N \rangle^2\right]\;,
\label{ent-lover-bound}
\end{equation}
where $N=\sum_{i=1}^{N_A}c_{i}^{\dagger} c_i$ is the particle number operator in the subsystem.
This lower bound was used to prove the behaviour of the entanglement entropy of multidimensional free fermions \cite{farkas-zimboras-2007} as well as of fractal-shaped partitions
\cite{fractal-boundary-real}.
The particle number fluctuations in the case of a random partition of a uniform probability are presented in Appendix \ref{deriv-of-particle-number-fluctuations} in one and two dimensions. From these we obtain the lower bounds
\begin{align}
\langle S \rangle_{\textrm{1D}} \ge &  \ln (2) L  p (1-p) + 4 \ln(2) \frac{p^2}{\pi^2} \ln (L-1) \label{1d-S-lb}\;, \\
 S \rangle_{\textrm{2D}} \ge & \ln (2) L^2 p(1-p) + 4  \ln(2) \frac{2 p^2}{\pi^2} L \left( \ln{L} - \frac{4}{\pi^2} \right) \;. \label{2d-S-lb}
\end{align}
These bounds are plotted in Fig. \ref{fig_1} and Fig. \ref{fig_6}. 
The leading term of the 2D lower bound in Eq. \ref{2d-S-lb} is proportional to the area of $2D$ percolation clusters 
\cite{grinchuk2003}, given by $E^{\textrm{tot}}_s = 4 L^2 p (1-p)$.

We note, that by shifting the parabola $x(1-x)$ upwards, one can also obtain upper bounds, for example with a shift of $0.08$
it holds as
\begin{equation}
s(x) \le  4\ln (2)  x(1-x) + 0.08\;.
\label{ineq-S-B}
\end{equation}
This leads to an upper bound for the prefactor of the volume term. 

\section{Limiting behaviours for $p \ll 1$ and for $1-p \ll 1$}
\label{sec:series-exp}
The average entanglement entropy can be calculated as a series expansion in $p$, performed in Appendix \ref{details-of-series-exp-2} for $1D$, leading to the following result up to ${\cal O}(p^3)$
\begin{equation}
\langle S \rangle_{\textrm{1D}} = (p \ln 2 - \alpha p^2 + \dots) L + \left( \frac{2 p^2}{\pi^2}  + \dots \right) \ln L\;,  
\label{1d-series-res}
\end{equation}
where $\alpha=0.5335$ is defined in Eq. (\ref{alpha}).

Similarly in two dimensions, the leading terms are
\begin{equation}
\langle S \rangle_{\textrm{2D}} = \left(p \ln 2 + {\cal O}(p^2)\right) L^2 +  \left(\frac{2 p^2}{\pi^2} + {\cal O}(p^4)\right) L \ln L \;.
\label{2d-series-res}
\end{equation}
The other limiting case, $1-p \ll 1$ can be treated as follows. Here, the logarithmic corrections approach the 
clean system's results, but due to dilution a volume term will appear. Let us now concentrate on the infinite subsystem (denoted by $B$), which - in the limit $1-p \ll 1$ - consists of two half lines (in $1D$) or
the whole plane without the square (in $2D$), as well as some isolated points from the interval (square). By neglecting correlations between isolated points, the correlation matrix of the infinite subsystem becomes block diagonal. One block contains the correlation matrix of the infinite subsystem  without the isolated sites, i.e. with $p=1$. The other block corresponds to the isolated sites, and has the dimension of the number of isolated points,
$\tilde{N}_{D}= (1-p) L^D$, containing $1/2$ in its diagonal as
\begin{equation}
C_{\textrm{B}} = \left[ \begin{array}{cc}
C(p=1) & 0 \\
0  &   \frac{1}{2} \mathbb{1}
\end{array} \right]
\end{equation}  
From this follows that the leading correction to the entanglement entropy is
\begin{equation}
\langle S \rangle = \Tr \, s(C_{\textrm{B}}) \approx S(p=1) + \ln(2) \, (1-p) L^d + {\cal O}\left[(1-p)^2\right]\;.
\label{series-1p}
\end{equation}
Note, that the series expansion results for the volume term agree with the lower bound for $p \ll 1$ and $1-p \ll 1$. 

\section{Numerical results}
\label{sec:numeric}
For finite subsystems of linear size, $L$, we have calculated the average entanglement entropy numerically. For the averaging process over different samples we have used three different methods. In the \textit{direct method} a large number of samples are generated at a fixed value of $p$ and the entanglement entropy is calculated for each sample, and this calculation is repeated for several values of $p$. This method generally gives an accurate average value at a given $p$, if the number samples is large enough ($10^6 \dots 10^8$) for each $p$, but comparing the results for different values of $p$ leads to large errors since the samples are different at each $p$.

In the so called \textit{indirect method} we use the relation
\begin{equation}
 \langle S \rangle(p)= \sum_{n=0}^N \langle S_n \rangle \left( \begin{array}{c}
                              N \\
                              n
                             \end{array}  \right) 
                          p^n (1-p)^{N-n} \;,  
                          \label{eq:indirect_method}
\end{equation}
where $N=L$ ($N=L^2$) in $1D$ ($2D$), and $\langle S_n \rangle$ denotes the average entanglement entropy of a subsystem of $n$ sites, where the averaging is performed over all possible partitions. In practice, we have generated a large number of random samples with uniform probability for all values of $n$. These samples and their entanglement entropy are stored and used to calculate $\langle S \rangle(p)$ for different values of $p$. The advantage of the indirect method is that we need comparatively less samples ($\sim 10^6$) and the numerical derivation with respect to $p$ is more smooth, compared to the direct method.

In our third, \textit{replica method} we generated for each random subsystem sample one (four) replicas in $1D$ ($2D$) and fused them together. By comparing the entanglement entropy of the original and the replicated sample one can cancel the leading volume term and gain direct access to the more interesting, subleading corrections. For the best results, our calculations have generally combined the replica method with the indirect method.

\subsection{$1D$ hopping model}

For the $1D$ hopping model we used the indirect method to calculate the average entanglement entropy for domain sizes $L=16, 32, \dots 1024$, with $10^6$ realizations in each case. In agreement with the analytical results on the lower bound in Eq.(\ref{1d-S-lb}) and the perturbation expansions in Eqs.(\ref{1d-series-res}) and (\ref{series-1p}), the leading contribution scales linearly with the number of sites in the domain, $L$.
%%%%%%%%%%%%%%%%%%%%%%%%% Fig 1 %%%%%%%%%%
\begin{figure}
\includegraphics[width=1.0\columnwidth]{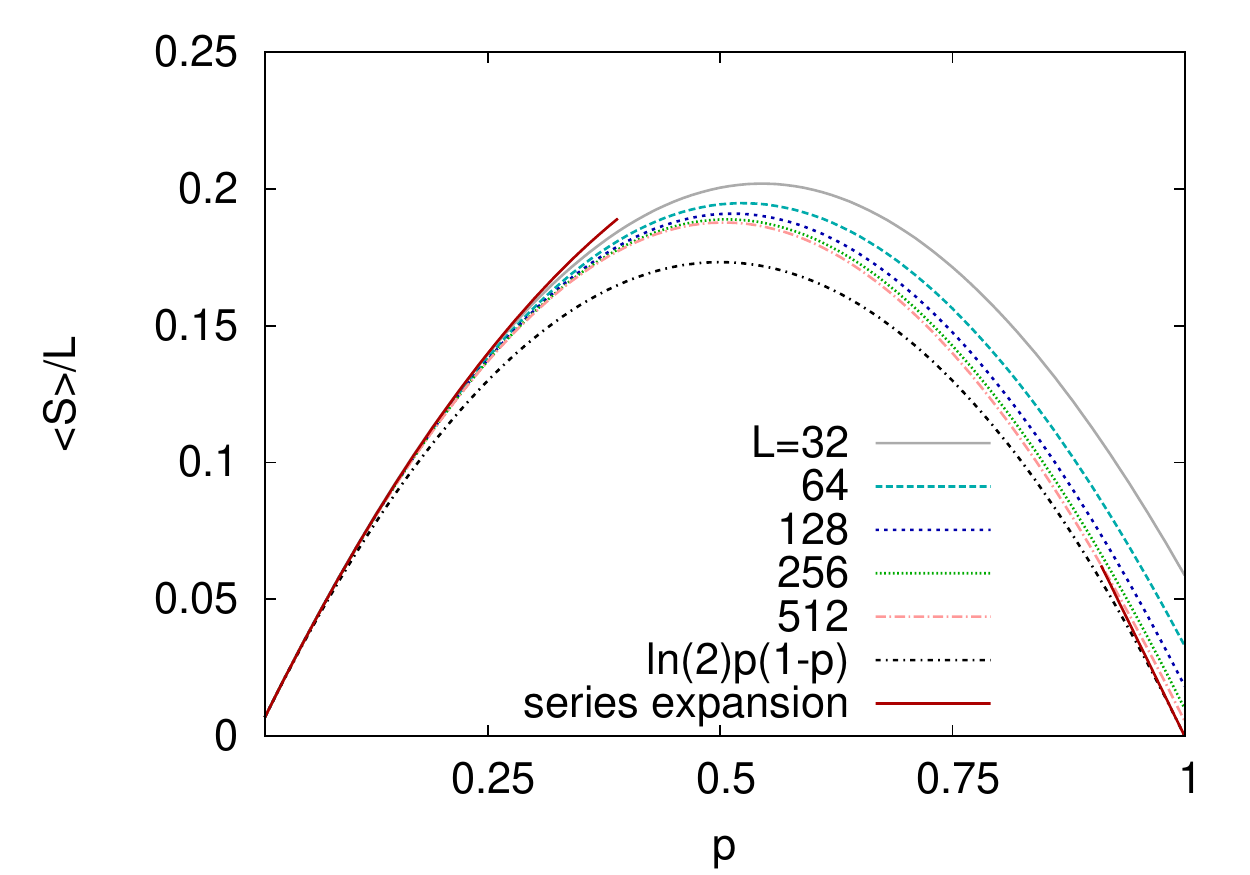}
\caption{Average entanglement entropy per domain size of the $1D$ hopping model calculated by the indirect method. The series expansion results for $p << 1$ and for $1-p<< 1$ are presented by red lines, whereas the obtained lower bound is drawn by a dotted line.\label{fig_1}}	
\end{figure} 
%%%%%%%%%%%%%%%%%%%%%
This is illustrated in Fig.\ref{fig_1}, where the average numerical vale of the entanglement entropy per domain size $\langle S \rangle/L$ is plotted together with the analytical results.

\begin{figure}
	\includegraphics[width=0.9\columnwidth]{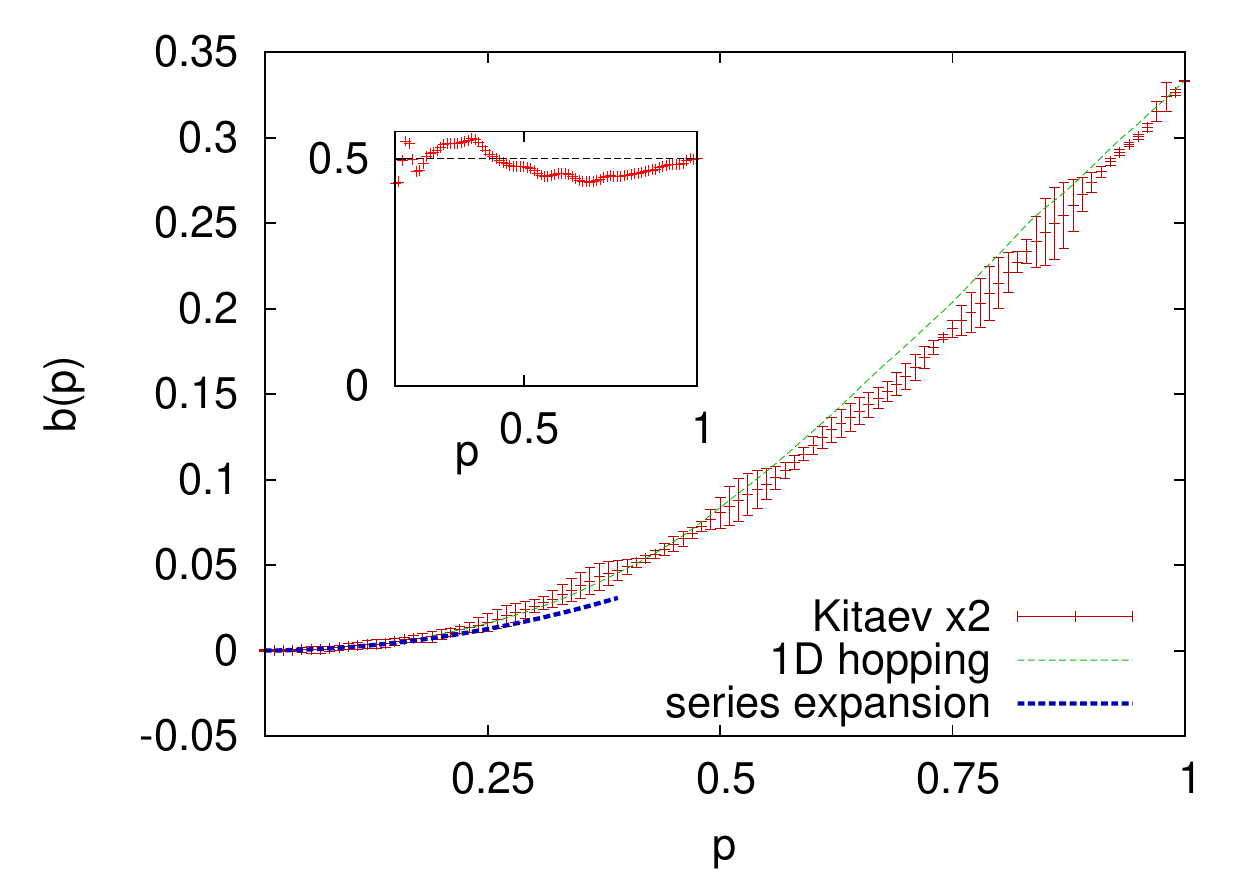}
    \caption{Prefactor of the logarithmic term of the average entanglement entropy of the $1D$ hopping chain (green) and twice the same the $1D$ critical Kitaev chain (red), calculated by the replica method, see text. The series expansion result for the $1D$ hopping chain at small $p$ is shown by a dotted line. In the inset the ratio of the two prefactors are shown and the dashed line represents the ratio of the conformal charges: $c^{\textrm{Kit}}/c^{\textrm{hop}}=1/2$.\label{fig_2}}
\end{figure}

\begin{figure}
	\includegraphics[width=0.9\columnwidth]{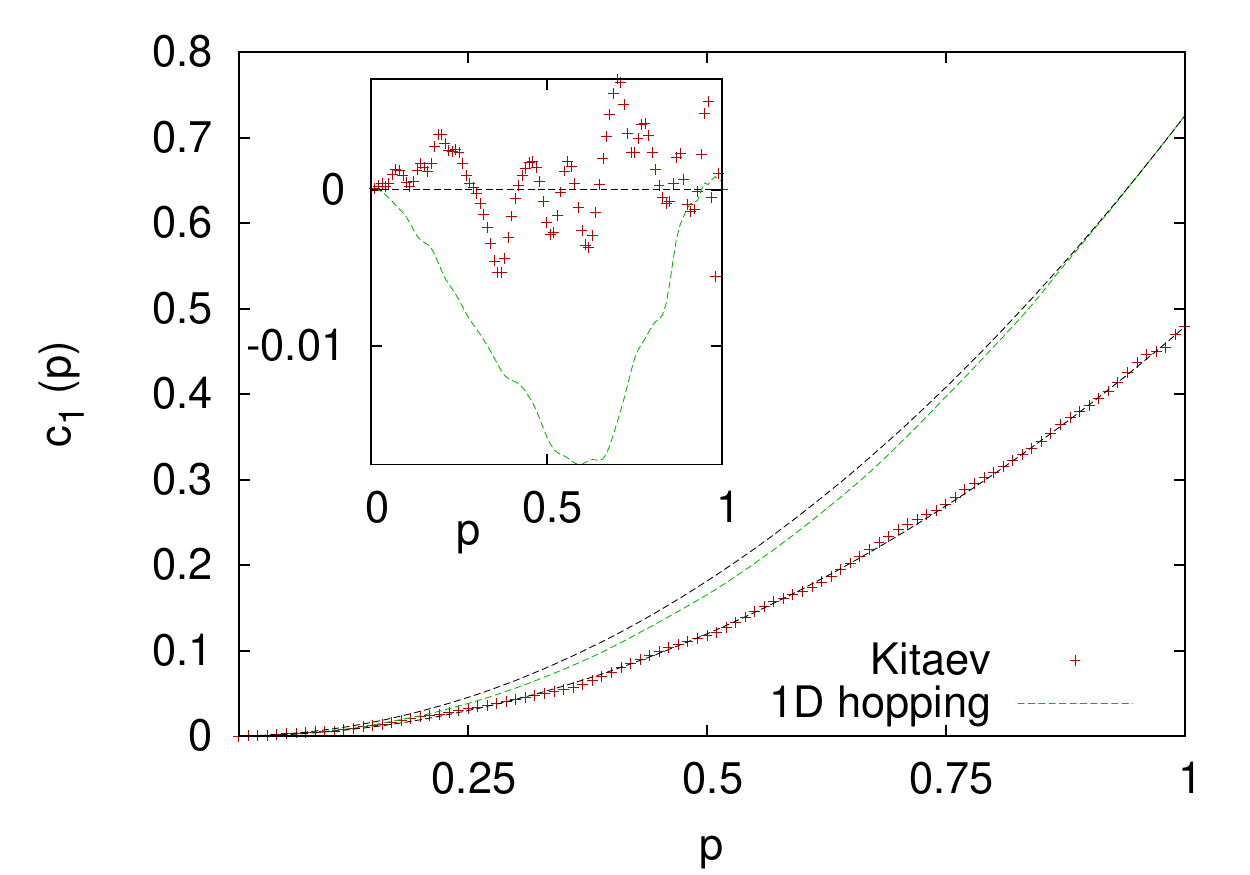}
\caption{The constant of the average entanglement entropy of the $1D$ hopping chain (green) and that of the $1D$ critical Kitaev chain (red), calculated by the replica method, see text. The dashed curves on the main panel corresponds to $c_1(1) p^2$ In the inset the difference $c_1(p)-c_1(1) p^2$ are shown for the two models. \label{fig_3}}
\end{figure}

The asymptotic behavior of the average entanglement entropy is expected to contain sub-leading terms in the form
\begin{equation}
 \langle S \rangle (L)=a(p) L + b(p) \ln L + c_1(p) \;.
 \label{num_fit_1D} 
\end{equation}
The prefactor of the volume contribution, $a(p)$, which can be represented by extrapolating the curves in Fig.\ref{fig_1}, is close to symmetric, $a(p) \approx a(1-p)$. More interesting are the subleading terms in Eq.(\ref{num_fit_1D}), which are conveniently analysed by the replica method using the difference
\begin{equation}
2\langle S \rangle(L,p)-\langle S \rangle_{\textrm{repl}}(2L,p)=b(p) (\ln L- \ln 2) +c_1(p)\;.
\label{1D-repl}
\end{equation} 
Here $\langle S \rangle_{\textrm{repl}}(2L,p)$ denotes the average entanglement entropy in the replicated samples, which are obtained by joining the same sample behind another copy. By comparing results at sizes $L$ and $2L$, finite-size estimates are calculated for the prefactor, $b(p)$, and the constant, $c_1(p)$, which are then extrapolated. These are plotted in Figs.\ref{fig_2} and \ref{fig_3}, respectively. 
%%%%%%%%%%%%%%%%%%%%%%%%% Fig 2 %%%%%%%%%%
%%%%%%%%%%%%%%%%%%%%%
The prefactor, $b(p)$, starts quadratically for small $p$, in agreement with the series expansion in Eq.(\ref{1d-series-res}), while at $p=1$ reaches the conformal result: $b(1)=c^{\textrm{hop}}/3=1/3$, see in Eq.(\ref{pure-1D-ent}). The constant, $c_1(p)$, also appears to start quadratically at small $p$, while at $p=1$ reaches the known result, as quoted below Eq.(\ref{2D-corr-func}). Interestingly, we have found an overall quadratic dependence: $c_1(p) \approx p^2 c_1(1)$, as illustrated in the inset of Fig.\ref{fig_3}.
%%%%%%%%%%%%%%%%%%%%%%%%% Fig 3 %%%%%%%%%%
%%%%%%%%%%%%%%%%%%%%%

\begin{figure}
 \includegraphics[width=0.9\columnwidth, angle=0]{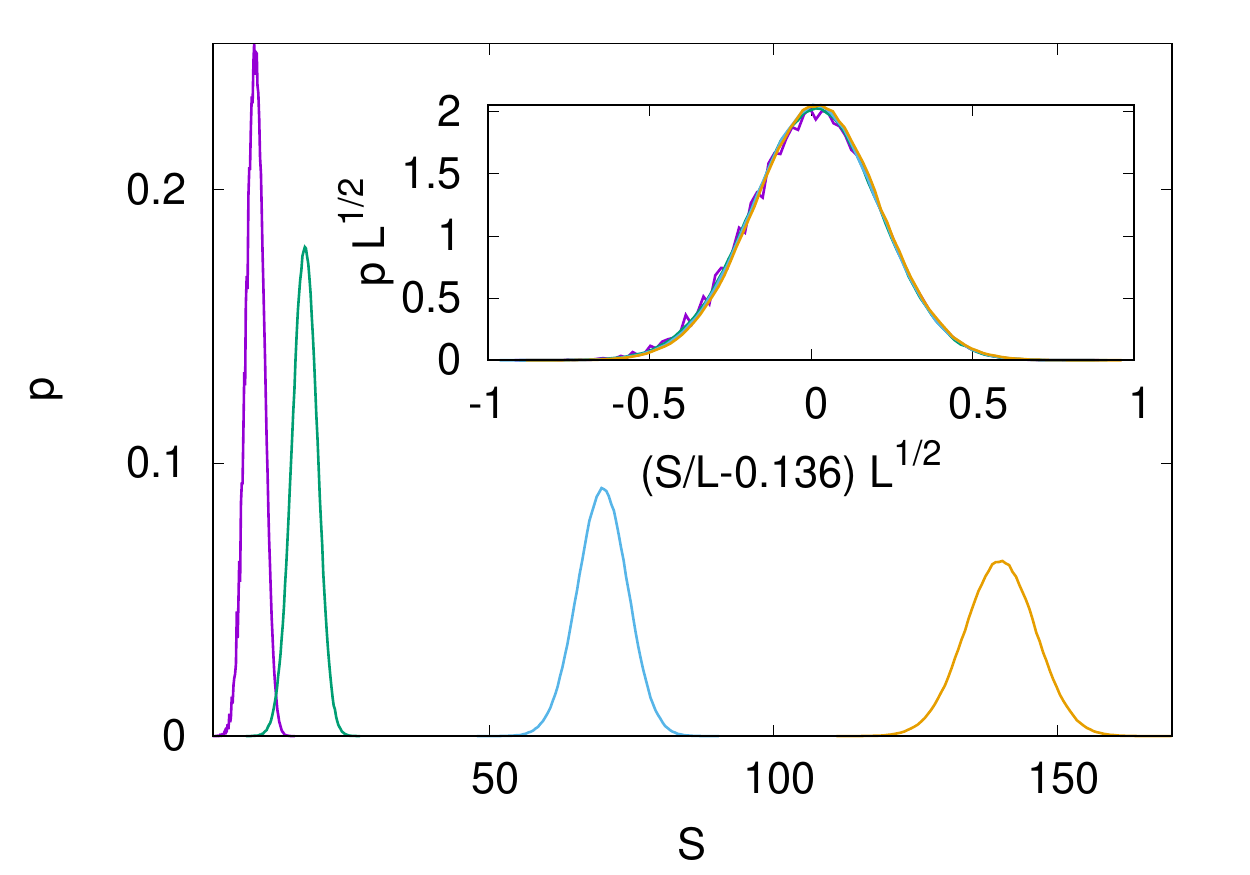}
 \caption{Probability distribution of the entanglement entropy of the $1D$ hopping chain at $p=0.25$ for different sizes: $L=64, \, 128, \, 512, \, 1024$, from left to right. In the inset the scaled curves are shown assuming Gaussian behaviour.
 \label{fig_4}}
\end{figure}

We have also studied the distribution of the entanglement entropy, shown in Fig. \ref{fig_4} for different sizes at $p=0.25$. These distributions are well represented by Gaussians, as illustrated in terms of scaled distributions in the inset. Similar, Gaussian distributions are observed for other values of $p$ as well, but close to $p=1$ there is a cross-over regime, where the volume contribution, $\ln (2) (1-p)L$, and the logarithmic term, $c/3 \ln L$, compete, see in Eq.(\ref{series-1p}).
%%%%%%%%%%%%%%%%%%%%%%%%% Fig 4 %%%%%%%%%%
%%%%%%%%%%%%%%%%%%%%%

\subsection{Critical Kitaev chain}

\begin{figure}
\includegraphics[width=0.9\columnwidth]{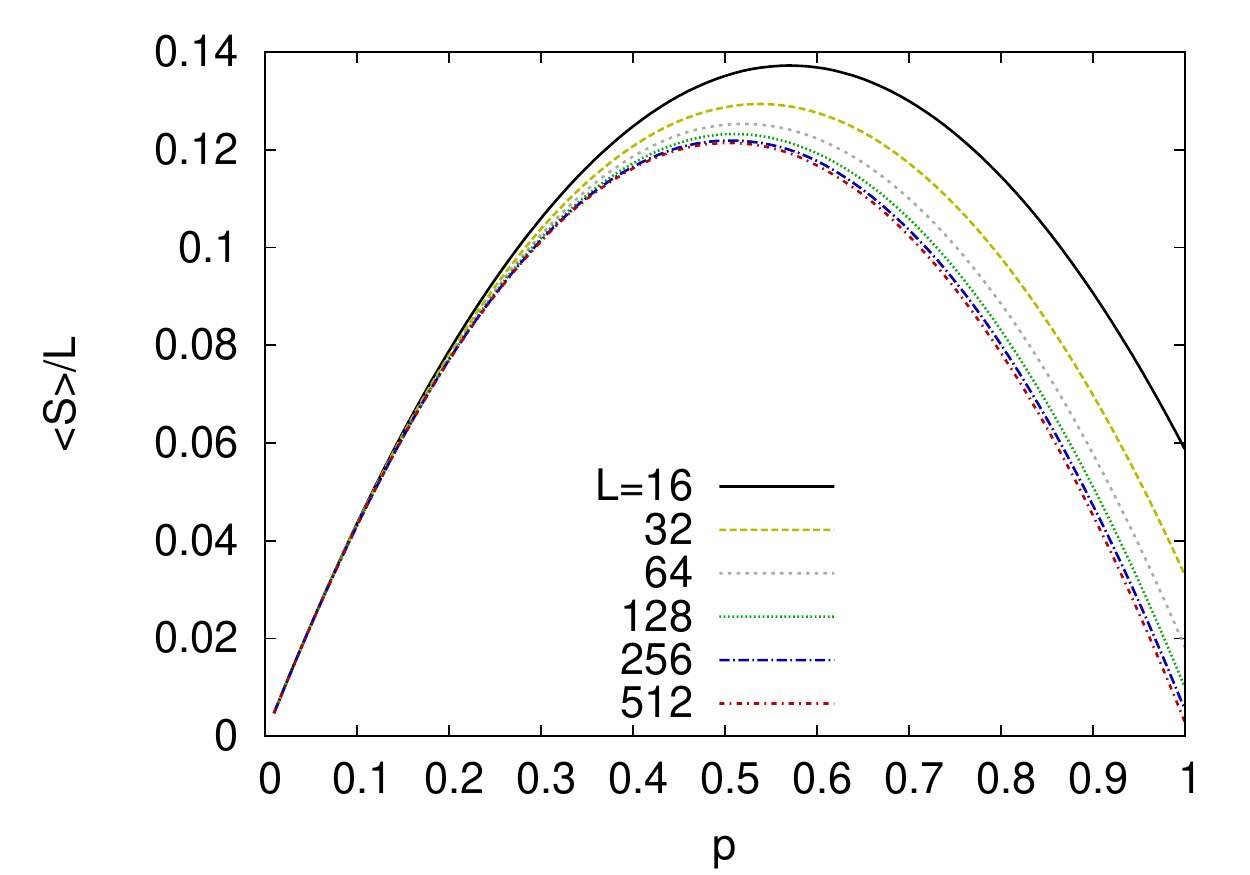}	
\caption{Average entanglement entropy per domain size of the critical Kitaev chain calculated by the indirect method. \label{fig_5}}
\end{figure}

For the critical Kitaev chain we have calculated the entanglement entropy of random partitions, as described in the previous subsection. Here we have used finite domains of size: $L=16,32,\dots,512$ and the averages are calculated by the indirect method over $10^6$ samples. As for the $1D$ hopping chain, the dominant contribution to the average entanglement entropy is the volume term, as illustrated in Fig.\ref{fig_5} where the average entanglement entropy per domain size is shown for different values of $L$. The shape of the extrapolated curve is similar to that of the $1D$ hopping chain in Fig.\ref{fig_1} and it is again approximately symmetric with respect to $p \to (1-p).$
%%%%%%%%%%%%%%%%%%%%%%%%% Fig 5 %%%%%%%%%%
%%%%%%%%%%%%%%%%%%%%%

The subleading correction terms are found to be in the same form as given in Eq.(\ref{num_fit_1D}). Using the replica method and Eq.(\ref{1D-repl}) we have calculated estimates for the prefactor of the logarithmic term, $b(p)$, as well as of the constant, $c_1(p)$, and their extrapolated values are plotted in Fig.\ref{fig_2} and Fig.\ref{fig_3}, respectively. Considering the prefactor, $b(p)$, its form is very similar to that found for the $1D$ hopping chain:
their ratio is given by $b(p)^{\textrm{Kit}}/b(p)^{\textrm{hop}}\approx 1/2 = c{\textrm{Kit}}/c^{\textrm{hop}}$. This is illustrated in the inset of Fig.\ref{fig_2}. As seen in Fig.\ref{fig_3} the constant term, $c_1^{\textrm{Kit}}(p)$, has also an approximately quadratic dependence: $c_1^{\textrm{Kit}}(p) \approx p^2 c_1^{\textrm{Kit}}(1)$, as illustrated in the inset of Fig.\ref{fig_3}.

\subsection{$2D$ hopping model}

\begin{figure}
\includegraphics[width=1.0\columnwidth]{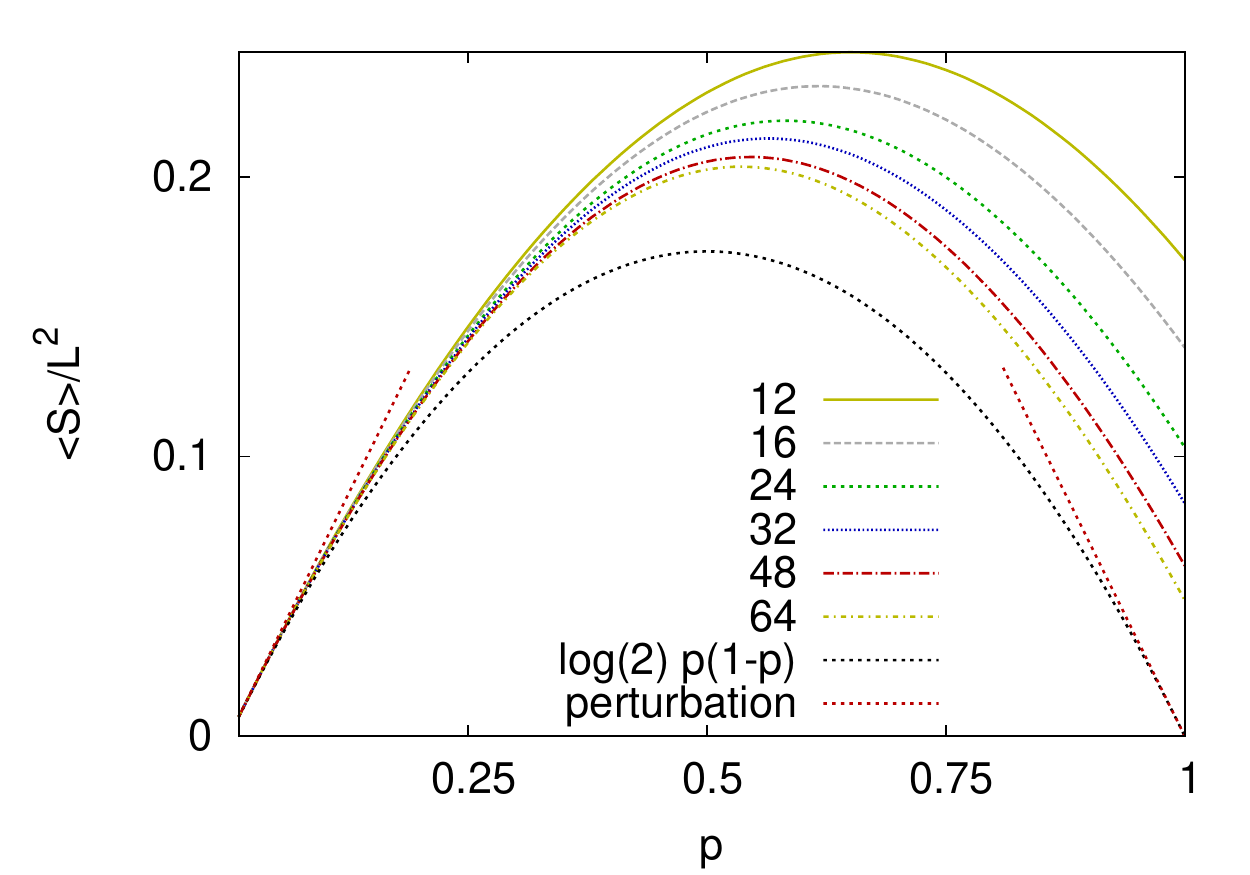}
\caption{Average entanglement entropy per domain volume of the $2D$ hopping model calculated by the indirect method. The perturbative result for $p \ll 1$ and $1-p \ll 1$ are presented by red dotted line, whereas the lower bound is drawn by a black dotted line.\label{fig_6}}	
\end{figure}

Here we consider the hopping model in a square lattice, in which the domain is an $L \times L$ square. We have calculated the entanglement entropy of samples having finite subsystems with linear extension $L=8,12,16,24,32,48$ and $64$, while averages are obtained through the indirect method over $10^5$ samples. According to the analytical results in Eqs.(\ref{2d-S-lb}) and (\ref{2d-series-res}) the average entanglement entropy is expected to be dominated by the surface term to which the first correction is logarithmic:
\begin{equation}
 \langle S \rangle(L)= a(p) L^2 + b(p) L \ln L + \dots\;.
 \label{2d_entropy}
\end{equation}
This is in agreement with our numerical results in Fig.\ref{fig_6}, showing the average entanglement entropy per domain surface. For increasing $L$, the curves approach the prefactor, $a(p)$, which is approximately symmetric, $a(p) \approx a(1-p)$. Comparing this figure with the one-dimensional results in Figs.\ref{fig_1} and \ref{fig_5} the convergence is here slower, due to considerably smaller linear size of the domains in $2D$.

%%%%%%%%%%%%%%%%%%%%%%%%% Fig 6 %%%%%%%%%%
%%%%%%%%%%%%%%%%%%%%%

\begin{figure}
\includegraphics[width=0.9\columnwidth, angle=0]{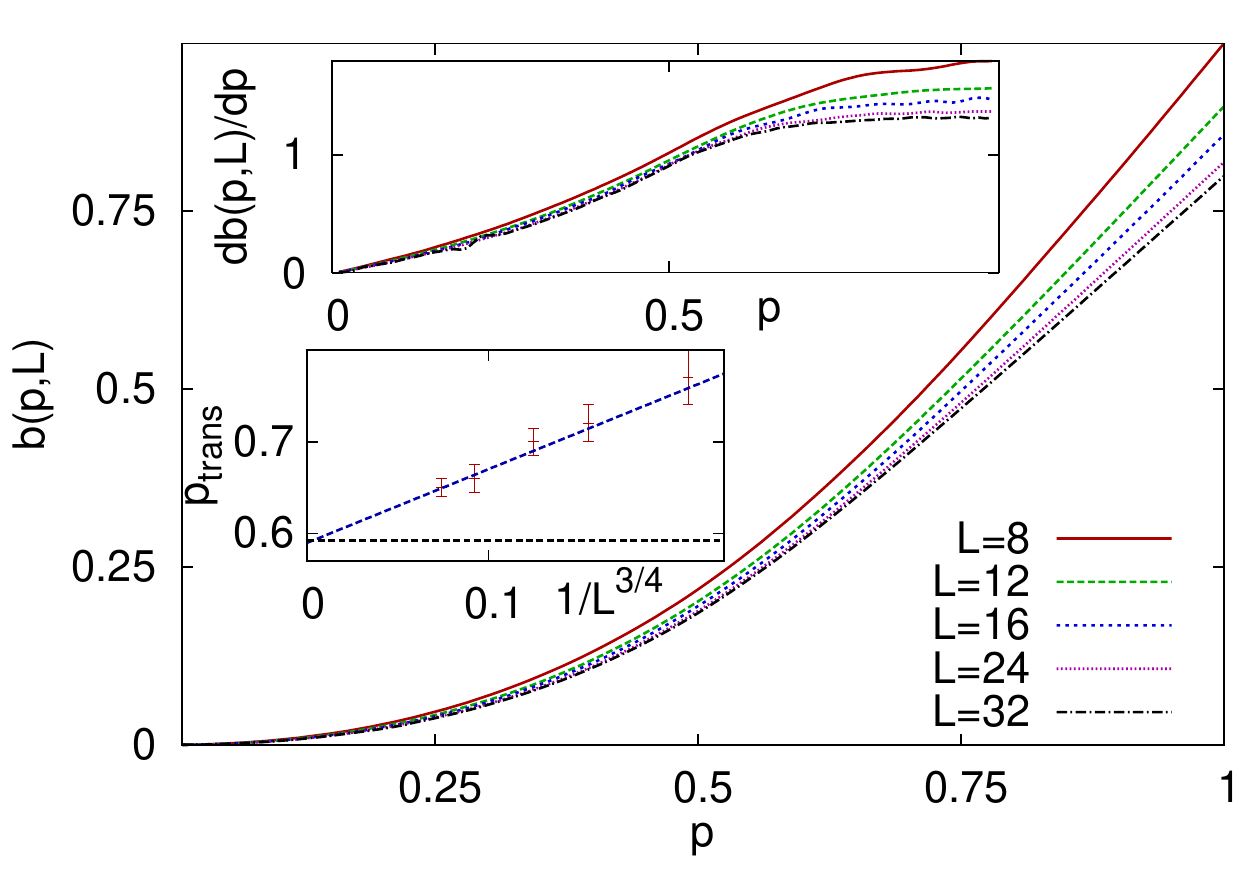}
\caption{Effective, size-dependent prefactor of the logarithmic correction term in the average entanglement entropy of the $2D$ hopping model calculated through the replica method in Eq.(\ref{2D_repl}). In the first inset the derivative $\displaystyle{\frac{\partial b(p,L)}{\partial p}}$ is shown. In the second inset the finite-size transition points are plotted as a function of $L^{-3/4}$, see the text. The dashed blue line is guide to the eye, the horizontal black dashed line represents the transition point for site percolation.\label{fig_7}}	
\end{figure}

The prefactor of the logarithmic term is estimated through the replica method: comparing the (four times) entanglement entropy of each sample of size $L$, with those composed of four joint identical samples, thus having a linear size $2L$
\be
4 \langle S \rangle(L,p)-\langle S \rangle_{\textrm{repl}}(2L,p)=b(p,L) 2L \ln(L/2)\;.
\label{2D_repl}
\ee
Eq.(\ref{2D_repl}) defines an effective, size-dependent prefactor, $b(p,L)$, which is plotted in Fig.\ref{fig_7}. As seen in this figure $b(p,L)$ starts quadratically for small $p$ and becomes approximately linear for larger values of the probability. To study this behaviour further we have calculated the derivative of $b(p,L)$ with respect to $p$, which is shown in the first inset of Fig.\ref{fig_7}. We note, that in the indirect method the differentiation of Eq.(\ref{eq:indirect_method}) can be performed at each value of $p$, which reduces the error of the calculation. Inspecting the behaviour of $\displaystyle{\frac{\partial b(p,L)}{\partial p}}$ we can identify two regions. In the first regime the derivative continuously increases, while in the second regime it becomes approximately constant. In finite subsystems there is an extended cross-over region between the two regimes, which, however, shrinks with increasing $L$. 
%%%%%%%%%%%%%%%%%%%%%%%%% Fig 7 %%%%%%%%%%
%%%%%%%%%%%%%%%%%%%%%

We summarize these findings in the conjecture that the change in the behaviour of $b(p,L)$ is related to the percolation transition\cite{stauffer}, which takes place in the random partitioning at a critical value $p_c=0.592$, if $L \to \infty$. To further check this hypothesis,
we have defined finite-size transition points between the two regions as the crossing point, where the linear continuation of the curve starting from $p=.5$ for $p>0.5$ reaches the value of the constant measured at $p \lessapprox 1$. These finite-size transition points are plotted in the second inset of Fig.\ref{fig_7} as a function of $L^{-1/\nu}$, with $\nu=4/3$ being the correlation-length critical exponent of $2D$ percolation, governing finite-size effects\cite{stauffer}. Indeed the extrapolated value of the finite-size transition point agrees with $p_c$, within the error of the calculation. We have also checked that the volume term with $a(p)$ shows no sign of a singularity at any value of $p$.

\section{Discussion}
\label{sec:discussion}

We have studied the entanglement entropy of critical free-fermion models in one and two dimensions, when the sites of the subsystem were taken from a hypercubic domain of linear size $L$ randomly, with probability $p$. We have investigated the average entanglement entropy by calculating lower bounds, by series expansions and performing extensive numerical calculations. When the entire system has infinite extent, the average entanglement entropy for $0<p<1$ is found to be dominated by the volume term $a(p) L^D$, which is supplemented by logarithmic corrections as $b(p) L^{D-1}\ln L$. The volume term is non-universal, which is connected to the fact, that the distribution of the entanglement entropy is Gaussian. On the contrary, the logarithmic correction is found to contain information about the universal, critical characteristics of the system.
In $1D$, comparing the results of the hopping chain and that of the Kitaev chain the prefactor of the logarithm is found to scale as: $b(p)=c f(p)$, where $f(p)$ is a universal, model independent function and $c$ is the central charge of the critical model. Interestingly, for both models the constant term is obtained in a pure quadratic form: $c_1(p) \approx p^2 c_1(1)$. In $2D$, for the hopping model $b(p)$ is shown to change its behaviour at the percolation transition point, $p_c$, where the random subsystem develops an infinite cluster. According to our numerical results the derivative, $\displaystyle{\frac{\partial b(p,L)}{\partial p}}$, is increasing with $p$ for $p<p_c$, but it saturates to a constant for $p>p_c$. This conjectured behaviour would be interesting to justify independently by physical argumentaarguments, perhaps even with some rigorous method.

Our study can be extended to several further directions as discussed next. 
\subsection{Finite environment}
\begin{figure}
\includegraphics[width=0.9\columnwidth, angle=0]{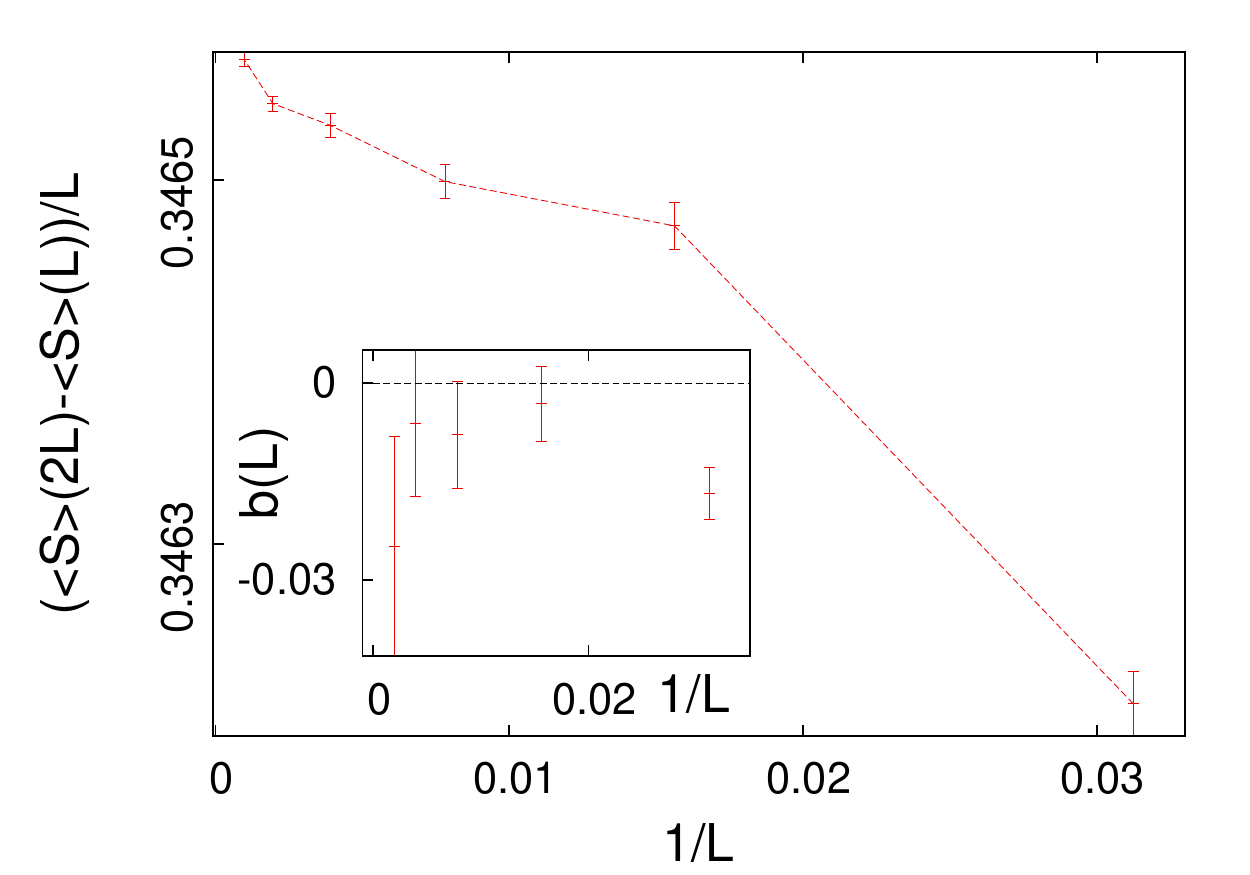}
\caption{
	%Upper panel: 
The ratio in Eq.(\ref{num_fit_1D_B}) versus $1/L$ in the case with $L/L_{\textrm{tot}}=1$ and $p=0.5$. In the inset the slope of the curve calculated by two point fit is presented, which defines finite-size estimates for the prefactor $b$. 
\label{fig_8}}
\end{figure}

We studied the case when the size of the entire system is $L_{\textrm{tot}} \to \infty$. For a finite value of $L_{\textrm{tot}}$, the results should depend on the ratio $L/L_{\textrm{tot}}$. In $1D$, for non-random partitions with $p=1$, the functional form of the entanglement entropy as a function of $L/L_{\textrm{tot}}$ is known from conformal invariance\cite{calabrese2004}. For the $1D$ hopping model with random partitions, we have checked that the prefactor $b(p)$ vanishes in the case $L/L_{\textrm{tot}}=1$. This is illustrated in Fig.\ref{fig_8}, where the ratio
\begin{equation}
 \frac{\langle S \rangle (2L)-\langle S \rangle (L)}{L \ln 2}=\frac{a}{\ln 2} + b \frac{1}{L} \;.
 \label{num_fit_1D_B}
\end{equation}
is plotted against $1/L$ at $p=1/2$. The slope of the points, which defines $b$, indeed tends to zero for $L \to \infty$, as seen in the inset of Fig.\ref{fig_8}.
%%%%%%%%%%%%%%%%%%%%%%%%% Fig 8 %%%%%%%%%%
%%%%%%%%%%%%%%%%%%%%%
\subsection{Position dependent selection probability}
Another potential extension of our study is to consider a different type of probability distribution for selecting the points of the subsystem. For the $1D$ hopping model, we have also checked a position dependent probability
\begin{equation}
 p_i=1-\frac{1}{2}\frac{1}{l_i^{\kappa}}\;,
\end{equation}
where $l_i=\textrm{min}(i,L-i)$ is the distance of the point $i$ from the nearest edge of the domain. By varying the decay exponent $\kappa \ge 0$ one can interpolate between the non-random partitioning with $p_i=1$ for $\kappa \to \infty$ and the uniform probability partitioning with $p_i=1/2$ for $\kappa=0$. The number of internal contact points between the subsystem and the environment scales as $\int_0^L l^{-\kappa} \textrm{d} l$, which is finite for $\kappa>1$, it scales as $\ln L$ for $\kappa=1$ and behaves as $L^{1-\kappa}$ for $0 \le \kappa < 1$.

\begin{figure}
\includegraphics[width=0.9\columnwidth, angle=0]{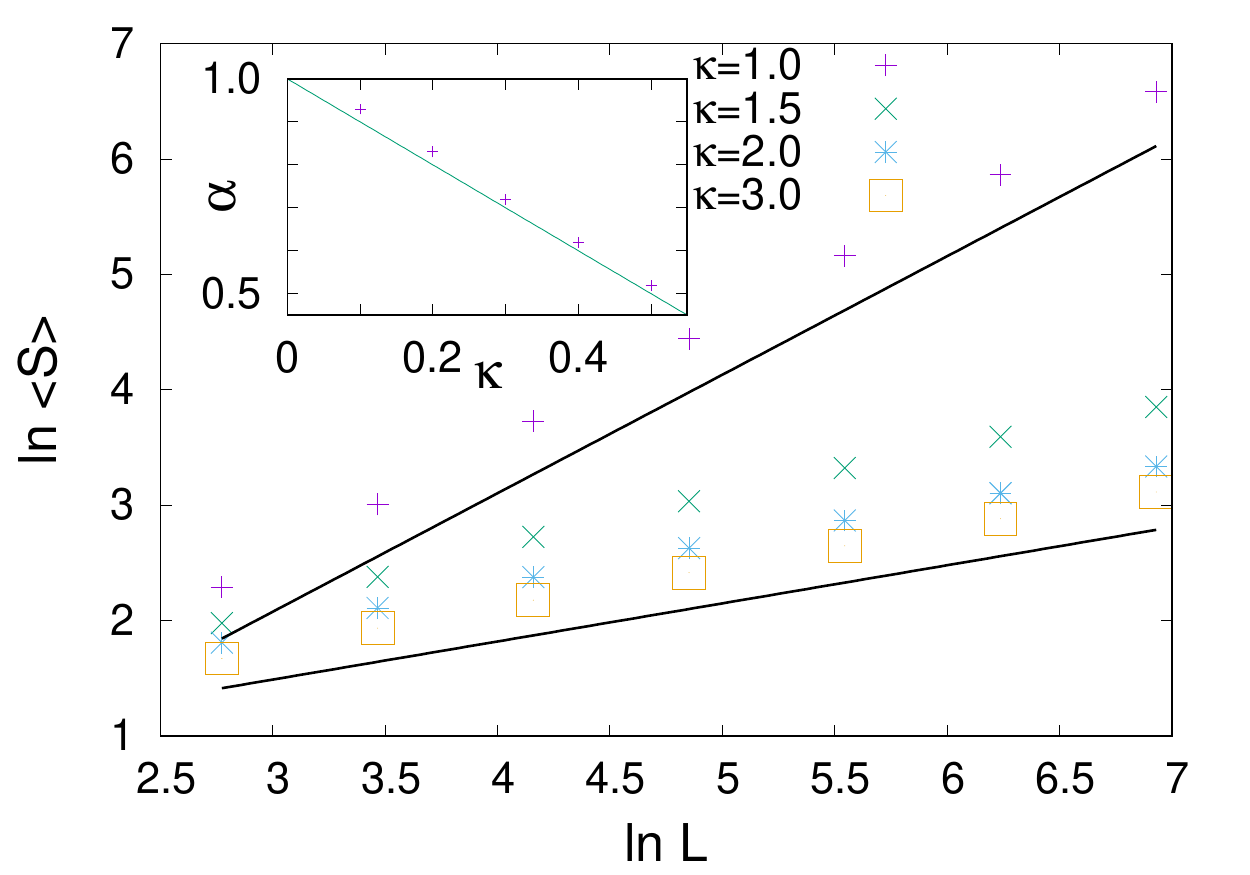}
 \caption{Average entanglement entropy with site dependent probability in one dimension. 
 The logarithm of the average entanglement entropy $\ln \langle S \rangle$ as a function of $\ln L$, for $1 \le \kappa \le 3$.  The black lines are guides to the eye,
 the slope  of the lower and upper lines are $1/3$ and $1/3+\ln(2)$, respectively. 
 In the inset the $\alpha$ exponent of the power-law increase of the average entanglement entropy is shown as a function of  $0.1 \le \kappa \le 0.5$, see text.
 \label{fig_9}}
\end{figure}

We have calculated the average entanglement entropy for different values of $\kappa$, ranging between $0.1$ and $3.0$ and the results are presented in Fig.\ref{fig_9}. For $\kappa > 1$ the dominant contribution is $\langle S \rangle = 1/3 \ln L$ as for the non-random partitioning case with $\kappa \to \infty$. This is due to the fact, that the "volume term", which scales with the number of contact points is now ${\cal O}(1)$, thus it is subleading. In the borderline case, $\kappa=1$, the size-dependence of $\langle S \rangle$ is still logarithmic, however with a different prefactor: $\langle S \rangle \approx (1/3 + \ln 2) \ln L$. The increase of the prefactor now is due to the "volume contribution", which scales also logarithmically. Finally, for 
$0 \le \kappa < 1$ the average entanglement entropy scales as: $\langle S \rangle \sim L^{\alpha}$, with $\alpha=1-\kappa$, which involves the number of contact points and now it is the dominant contribution. These results are illustrated in the inset of Fig.\ref{fig_9} .

%%%%%%%%%%%%%%%%%%%%%%%%% Fig 9 %%%%%%%%%%
%%%%%%%%%%%%%%%%%%%%%

\subsection{Some open problems}

It would be interesting to try to determine the function $f(p)$ by a different method, perhaps even analytically. It could also be interesting to check, if the observed universality holds for non-fermionic models, too. In addition, one can consider models with random couplings and pose similar questions treated in this paper.

\begin{acknowledgments}
This work was supported by the National Research Fund under Grants No. K128989, No. K115959 and No. KKP-126749 and by the Deut\-sche For\-schungs\-ge\-mein\-schaft through the Cluster of Excellence on Complexity and Topology in Quantum Matter ct.qmat (EXC 2147). This publication was made possible through the support of a grant from the John Templeton Foundation. The opinions expressed in this publication are those of the authors and do not necessarily reflect the views of the John Templeton Foundation.
We thank R\'obert Juh\'asz, Carsten Timm and Zolt\'an Zimbor\'as for helpful discussions. 
\end{acknowledgments}

\appendix
\section{Particle number fluctuations}
\label{deriv-of-particle-number-fluctuations}
In $1D$, the average particle number fluctuations are given by
\beqn
&&  \left[ \langle N^2 \rangle - \langle N \rangle^2\right]_{1D}=\langle \textnormal{Tr}~C \rangle-\langle \textnormal{Tr}~C^2 \rangle
\nonumber \\
&=&\left\langle \sum_{i \in A}c^{\dagger}_i  c_i\right\rangle
-\left\langle \sum_{i \in A} \sum_{j \in A}|C(i,j)|^2\right\rangle\nonumber \\
&=&\frac{1}{2} p L-p\sum_{i=1}^L |C(i,i)|^2-p^2 \sum_{i \ne j}|C(i,j)|^2\nonumber \\
&=&\frac{1}{2} p L-\frac{1}{4} p L-2p^2\sum_{k=1}^{L/2}(L-2k+1) \displaystyle{\frac{1}{\pi^2 (2k-1)^2}}\nonumber \\
&\approx&\frac{1}{4} L p(1-p) + \displaystyle{\frac{p^2}{\pi^2}} \ln L \;.
\eeqn
In $2D$, we obtain by a similar calculation
\begin{equation}
  \left[ \langle N^2 \rangle - \langle N \rangle^2  \right]_{2D} \approx  \frac{L^2}{4} p(1-p) + \frac{2 p^2}{\pi^2} L \left( \ln(L) - \frac{4}{\pi^2} \right) \;.
\end{equation}

\section{Series expansion}
\label{details-of-series-exp-2}
In $1D$, the average entanglement entropy is written as
\begin{equation}
\langle S \rangle = \sum_{N=1}^L p^N (1-p)^{L-N} \sum_{A, |A|=N} \Tr [s(C_A)]
\label{ser_exp_p}
\end{equation}
where the second sum goes for every subsystems $A$ including $N$ sites, and having the correlation matrix $C_A$.
For small $p$, we keep in Eq.(\ref{ser_exp_p}) the terms with $N=1$ and $N=2$ and omit terms with ${\cal O}(p^3)$, leading to
\begin{equation}
\langle S \rangle =pL\ln 2 - p^2(L-1)L\ln 2+ p^2 \sum_{A, |A|=2} \Tr [s(C_A)]+{\cal O}(p^3)
\label{ser_exp_p1}
\end{equation}
The correlation matrix of the two site subsystem is
\begin{equation}
C= \left[ \begin{array}{cc}
1/2 & C(m,n) \\
C(m,n) & 1/2 
\end{array} \right]
\end{equation}
where $m$ and $n$ are indices of the points included in the subsystem and $C(m,n)$ is given in Eq.(\ref{1D-corr-func}). The eigenvalues of $C$ are
\be
\zeta_{n,m}=
\begin{cases}
1/2 &\textrm{if}  \quad |n-m|=\textrm{even} \\
1/2 \pm \displaystyle{\frac{1}{\pi(n-m)}} &\textrm{if} \quad |n-m|=\textrm{odd}\;.
\end{cases}
\ee
The entropy contribution from a term with $|n-m|=\textrm{even}$ is $2 \ln 2$, whereas from a term with $|n-m|=\textrm{odd}$ is given by $2 \ln 2 + \Delta(n,m)$, with
\be
\Delta(n,m)= \left[2s\left(1/2 + \frac{1}{\pi(n-m)}\right) - 2\ln 2\right]\;.
\ee
Substituting this into Eq.(\ref{ser_exp_p1}) leads to
\beqn
\langle S \rangle &=&pL\ln 2+p^2 \sum_{n,m=1; \; |n-m| \, \textrm{odd} }^{L}\Delta(n,m)\nonumber \\
&=&pL\ln 2 +np^2 \sum_{k=1}^{L/2} (L-2k+1)\Delta(2k,1)\nonumber \\
&&\approx L (p \ln2  - \alpha p^2 )+p^2 \frac{2}{\pi^2} \ln L \;,
\eeqn
with
\begin{equation}
\alpha = -\sum_{k=1}^{\infty }  \left[ 2 s\left(1/2+\frac{1}{\pi (2k-1)}\right) - 2 \ln 2\ \right] \approx  0.5335\;.
\label{alpha}
\end{equation}

\end{document}